\documentclass{article}

\usepackage{PRIMEarxiv}
\usepackage[utf8]{inputenc} 
\usepackage[T1]{fontenc}    
\usepackage{hyperref}       
\usepackage{natbib}         
\usepackage{url}            
\usepackage{booktabs}       
\usepackage{amsfonts}       
\usepackage{amsmath}        
\usepackage{nicefrac}       
\usepackage{microtype}      
\usepackage{pdflscape}      
\usepackage{fancyhdr}       
\usepackage{graphicx}       
\usepackage{float}
\usepackage{enumitem}
\usepackage{todonotes}
\usepackage{caption}
\usepackage{graphicx}
\usepackage{subcaption}

\title{Disentangling Sampling from Training Budget in Class-Imbalanced CT Body Composition Segmentation}

\author{
    Iason Skylitsis \\
    Department of Biomedical Engineering \& Physics \\
    Amsterdam University Medical Center \\
    Amsterdam, The Netherlands \\
    Informatics Institute, Faculty of Science \\
    University of Amsterdam \\
    Amsterdam, The Netherlands \\
    \texttt{i.skylitsis@amsterdamumc.nl} \\
    \And
    Dimitrios Karkalousos \\
    Department of Biomedical Engineering \& Physics \\
    Amsterdam University Medical Center \\
    Amsterdam, The Netherlands \\
    Informatics Institute, Faculty of Science \\
    University of Amsterdam \\
    Amsterdam, The Netherlands \\
    \texttt{d.karkalousos@amsterdamumc.nl} \\
    \And
    Ivana Išgum \\
    Department of Radiology \\
    Mayo Clinic \\
    Rochester, United States \\
    Department of Biomedical Engineering \& Physics \\
    Amsterdam University Medical Center \\
    Amsterdam, The Netherlands \\
    Department of Radiology \\
    Amsterdam University Medical Center \\
    Amsterdam, The Netherlands \\
    \texttt{isgum.ivana@mayo.edu} \\
}

\begin{document}
\maketitle

\begin{abstract}
Class imbalance is a fundamental challenge in medical image segmentation, where frequent classes typically dominate training at the expense of rare classes.
Loss-based approaches mitigate imbalance by reweighting the per-pixel loss within the batch, while sampling strategies control which images enter the batch. Yet neither explicitly controls which classes appear within the batch, leaving rare-class exposure only partially rebalanced.
In this work, we adopt episodic sampling from few-shot learning to promote class-balanced batch construction in a fully supervised setting. We decouple episodic sampling from its conventional metric-learning context and evaluate it in body composition segmentation in CT.
We compare episodic sampling against random and weighted sampling on nine muscle and adipose tissues, derived from 210 scans of the public SAROS dataset. Training is performed under full- and low-data regimes, with additional comparisons under matched training iteration budgets.
Under full-data training, all three strategies performed comparably (mean Dice 0.882 for episodic, 0.878 for random and weighted). Under low-data training, episodic sampling outperformed random and weighted (0.787 vs. 0.758 and 0.762), driven by a 12-fold difference in training iterations. Under matched training budgets, random and weighted overfit earlier, while episodic improved for approximately three times more iterations before plateauing.
Our findings identify the training iteration budget as under-recognized confound in sampling strategies, motivating iteration-aware evaluation protocols for small datasets. Furthermore, the residual advantage of episodic sampling is consistent with an implicit regularization effect of class-balanced batches, offering a low-cost, model-agnostic strategy for class-imbalanced medical image segmentation. Code is available at \url{https://github.com/iasonsky/episodic-sampling}.
\end{abstract}


\section{Introduction}

Standard supervised learning typically samples training instances uniformly, implicitly assuming a balanced data distribution. In dense prediction tasks, such as semantic segmentation of medical images, this assumption rarely holds. Classes like background and large anatomical structures comprise orders of magnitude more pixels than small tissues or lesions. In addition, since segmentation models are trained by computing a loss over every pixel in each image, the gradient updates that drive learning are dominated by frequent classes, which contribute the majority of the per-pixel loss terms. As a result, rare classes receive proportionally fewer gradient updates, leading to models biased toward frequent classes, overfitting, and reduced segmentation accuracy under class imbalance.

Class imbalance in medical image segmentation is typically mitigated at the loss level. For example, weighted cross-entropy assigns higher penalties to underrepresented classes. Dice loss \citep{DBLP:journals/corr/SudreLVOC17} is inherently robust to class frequency disparities by optimizing for region overlap. Focal loss \citep{lin2018focallossdenseobject} down-weights easy examples to focus training on hard ones. In addition, compound losses, such as cross-entropy combined with Dice, have been shown to handle class imbalance more robustly than single losses~\citep{Ma2021}, establishing them as a standard practice~\citep{Isensee2020}. Despite their effectiveness on the gradient signal, loss-based approaches do not alter the training distribution itself.

Complementary to gradient-level mitigation, class imbalance can also be addressed at the input level by shaping batch composition through the sampling process. For example, standard weighted sampling assigns higher selection probabilities to images containing rare classes. More sophisticated approaches include oversampling and undersampling \citep{HaiboHe2009}, class-aware and repeat-factor sampling \citep{DBLP:journals/corr/abs-1908-03195,yaman2023instanceawarerepeatfactorsampling}, patch- and volume-level sampling weighted by class presence \citep{Kamnitsas2017}, and per-image imbalance-ratio weighting \citep{Roshan2024}. Despite this diversity, such methods control which images enter the batch without controlling class composition within it. Rare-class voxels therefore remain embedded in dominant-class context, so the per-voxel gradient signal remains only partially rebalanced.

Sampling has also been used for variance reduction, refining optimization by reducing gradient noise during batch construction. \citet{zhao2014stochastic} showed that stratified mini-batch sampling tightens convergence bounds relative to uniform sampling, requiring fewer iterations to reach a given error level. Subsequent work formalized this concept into importance sampling, a complementary variance-reduction tool for deep learning. \citet{DBLP:journals/corr/abs-1803-00942} and \citet{you2023rethinkingsemisupervisedmedicalimage} applied importance sampling to prioritize the most representative pixels within semantically similar groups. However, later work showed minimal effect of importance sampling on the asymptotic decision boundary of overparameterized networks \citep{byrd2019effectimportanceweightingdeep}, often underperforming fine-tuned baselines \citep{shwartzziv2023simplifyingneuralnetworktraining} or even compromising representation quality \citep{kang2020decoupling,zhou2020bbn}.

Across input-level rebalancing and stratified or importance sampling, existing methods adjust how often individual images are drawn into a batch, whether to compensate for class imbalance or to reduce gradient variance. The class composition within each batch, however, is not explicitly controlled. A notable exception comes from few-shot prototypical learning \citep{snell2017prototypicalnetworksfewshotlearning}, where training mini-batches (episodes) are sampled from a controlled subset of classes, with each episode containing a support and query set. Episodic sampling has shown promising results on imbalanced medical image segmentation \citep{Ouyang2020,guo_imbalanced_2025,roshan_deep_2024,tian_implicit-explicit_2024}, yet its mechanism is typically entangled with metric-based learning objectives. The episodic batch-construction logic, however, is independent of metric learning and model-agnostic, suggesting plug-and-play applications in fully supervised learning.

Nevertheless, adapting episodic sampling in supervised training, raises a methodological challenge that has received limited attention in medical image segmentation. Sampling-strategy comparisons typically specify training schedules in epochs, including learning rate milestones, early stopping patience, and maximum training duration, implicitly coupling the effective training iterations budget to dataset size. When samplers with different numbers of iterations per epoch are compared under such schedules, this coupling introduces a confound. Previous work in classification has shown that the apparent gains of specialized imbalanced sampling schemes can shrink substantially when iteration budgets are matched \citep{li2020budgetedtrainingrethinkingdeep,arazo2021importantimportancesamplingdeep}, or when compared against fine-tuned baselines \citep{shwartzziv2023simplifyingneuralnetworktraining}. The state-of-the-art nnU-Net framework \citep{Isensee2020} sidesteps this by setting a budget of $250{,}000$ training iterations regardless of dataset size, yet typically sampling schemes specify schedules in epochs.

In this work, we decouple the episodic batch construction from metric-based learning and apply episodic sampling in standard supervised training. We compare episodic sampling against standard random and weighted sampling under two training-data regimes: a full-data setting using all annotated volumes, and a low-data setting retaining 10\% via patient-level subsampling, which sharpens class underexposure. To isolate the contribution of the sampling mechanism from the training budget, we further evaluate the strategies under matched iteration budgets and examine their interaction with epoch-based scheduling. We focus on multi-class body composition segmentation in Computed Tomography (CT), an inherently class-imbalanced task where large adipose and muscle compartments coexist with small, spatially localized structures, yielding class frequencies that differ by several orders of magnitude within each scan. Beyond the methodological fit, our work targets fine-grained segmentation of multiple muscle structures, a setting under-explored by existing body composition analysis pipelines, which typically operate on a single 2D slice or collapse the problem to coarse tissue labels \citep{blankemeier_comp2comp_2023, hofmann_validation_2025}.

\section{Methods}

We investigate three sampling strategies for class-imbalanced body composition segmentation: random, weighted, and episodic. To isolate the effect of the sampling strategy, the network architecture, loss function, and optimization settings are held constant across all experiments. The following sections detail the dataset and the construction of reference annotations (Sec.~\ref{subsec:data}), the sampling strategies (Sec.~\ref{subsec:sampling}), the network architecture, training protocol, and evaluation metrics (Sec.~\ref{subsec:architecture}), and the experimental setup (Sec.~\ref{subsec:exp}).

\subsection{Data}
\label{subsec:data}

We used 210 CT scans from the publicly available Sparsely Annotated Region and Organ Segmentation (SAROS) dataset \citep{koitka_saros_2024}. SAROS comprises 900 CT scans curated from 28 collections within The Cancer Imaging Archive (TCIA), of which only 210 were freely available without additional licensing requirements. Our experiments were therefore restricted to this subset, the characteristics of which are summarized in Table~\ref{tab:dataset_overview}. Ethics approval for the use of these data was granted by the Medical Ethical Committee (METC) of Amsterdam UMC.

SAROS provides annotations for thirteen semantic body regions and six body-part labels, including subcutaneous adipose tissue (SAT) and skeletal muscle (SM). Reference annotations were created sparsely, by annotating every fifth axial slice. However, as noted in the dataset description, in the reference annotations the skin was merged into SAT and SM was segmented as a single contiguous structure rather than separated into individual muscles, with fascias and intermuscular adipose tissue (IMAT) incorporated into the muscle label. Additionally, after manual inspection we identified residual SM overestimation, including expansion into physiologically implausible regions (e.g., underneath the neural spine and along the vertebral surfaces), and remaining inclusion of fascia, scar tissue, or IMAT.

\begin{table}[htp]
\centering
\caption{Overview of the 210 publicly available scans from the SAROS dataset \citep{koitka_saros_2024}.}
\label{tab:dataset_overview}
\begin{tabular}{l c}
\toprule
\textbf{Characteristic} & \textbf{Value} \\
\midrule
\multicolumn{2}{l}{\textit{Demographics}} \\
\quad Sex & \\
\quad\quad Female / Male / Not reported & 124 (59.0\%) / 64 (30.5\%) / 22 (10.5\%) \\
\quad Age & \\
\quad\quad Mean $\pm$ SD (yrs) / Range / Not reported & $58.4 \pm 17$ / 2 mo\,--\,90 yr / 42 \\
\addlinespace
\multicolumn{2}{l}{\textit{Imaging type}} \\
\quad Contrast-Enhanced CT Single-Phase / Multi-Phase & 95 (45.2\%) / 60 (28.6\%) \\
\quad PET-CT & 50 (23.8\%) \\
\quad Non-contrast CT & 5 (2.4\%) \\
\addlinespace
\multicolumn{2}{l}{\textit{Scanner vendor}} \\
\quad Siemens & 114 (54.3\%) \\
\quad GE Medical Systems & 49 (23.3\%) \\
\quad Not reported & 47 (22.4\%) \\
\addlinespace
\multicolumn{2}{l}{\textit{Longitudinal temporal event}} \\
\quad Surgery & 86 (41.0\%) \\
\quad Diagnosis & 41 (19.5\%) \\
\quad Baseline & 30 (14.3\%) \\
\quad Not reported & 53 (25.2\%) \\
\bottomrule
\end{tabular}
\end{table}

To address these issues, we refined and expanded the existing labels with additional muscle and adipose tissue segmentations obtained from the Body-and-Organ Analysis (BOA) tool \citep{haubold2024boa}. Nine tissue classes were defined: erector spinae muscle (ESM), intermuscular adipose tissue (IMAT), pectoral muscle (PEM), psoas major (PSM), quadratus lumborum (QLM), rectus abdominis (RAM), subcutaneous adipose tissue (SAT), skeletal muscle (SM), and visceral adipose tissue (VAT). SM was defined as the residual region after exclusion of the five muscle subgroups. Hounsfield Unit (HU) thresholds were applied to constrain the masks to physiologically plausible attenuation ranges: $[-29, +150]$~HU for muscle tissues, $[-190, -30]$~HU for subcutaneous and intermuscular adipose tissue, and $[-150, -50]$~HU for visceral adipose tissue. VAT was further refined to prevent overlap with organs or bones by subtracting an organ-and-bone mask obtained with BOA, and isolated clusters of fewer than five voxels were removed via 3D connected-component analysis. IMAT was defined as the thresholded region within all muscle masks, with overlap with SAT and VAT explicitly subtracted. The refined labels were restricted to the anatomical trunk and extremities using the provided body-part annotations. 

\begin{figure}[htp]
    \centering
    \includegraphics[width=0.75\linewidth]{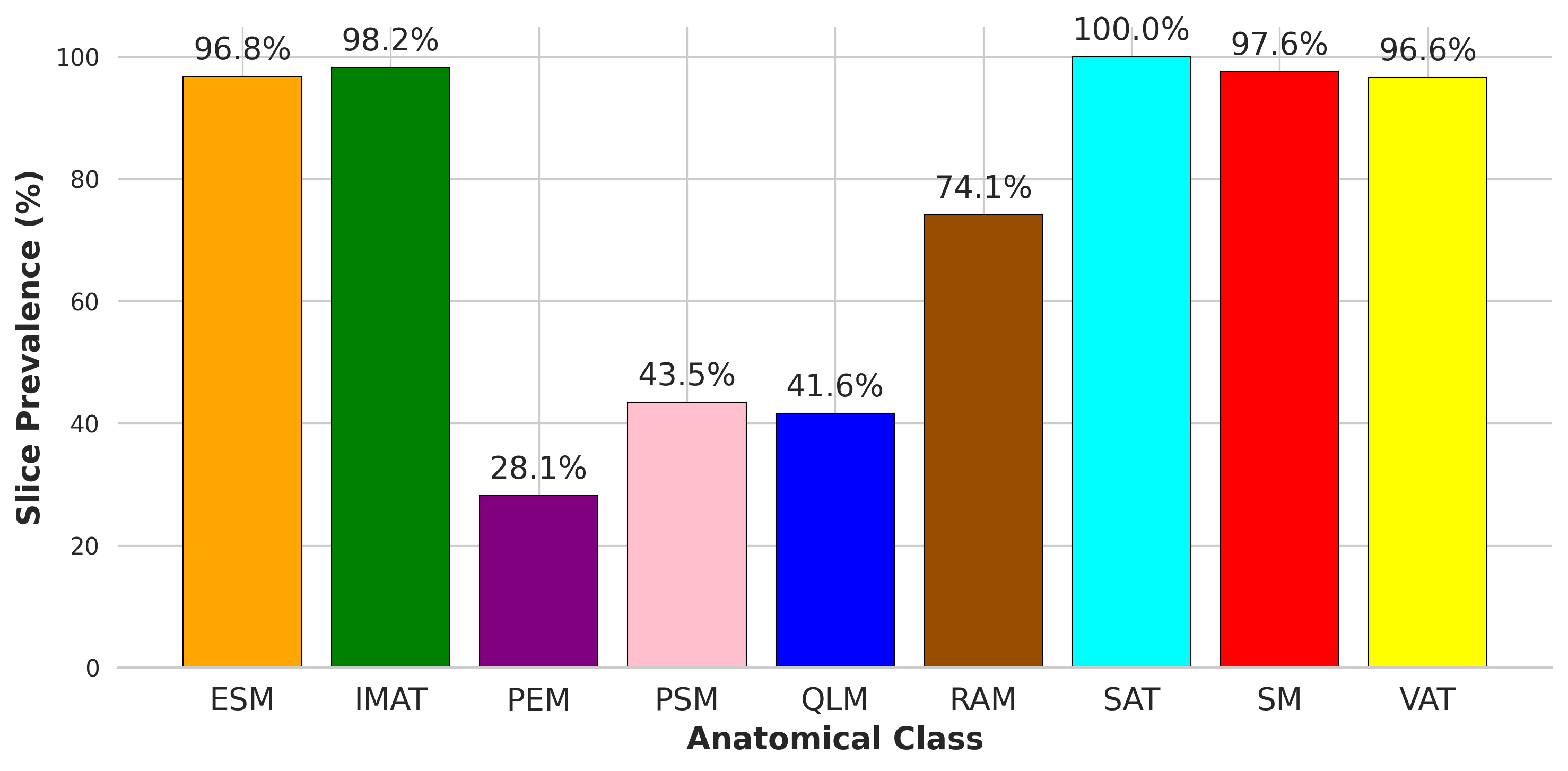}
    \caption{Slice-wise prevalence of the nine tissue classes. The y-axis indicates the percentage of slices in which each class is present. ESM: erector spinae muscle (orange); IMAT: inter-/intramuscular adipose tissue (green); PEM: pectoral muscle (purple); PSM: psoas muscle (pink); QLM: quadratus lumborum muscle (blue); RAM: rectus abdominis muscle (brown); SAT: subcutaneous adipose tissue (cyan); SM: skeletal muscle (red); VAT: visceral adipose tissue (yellow).}
    \label{fig:class_prevalence}
\end{figure}

\begin{figure}[htp]
    \centering
    \includegraphics[width=0.6\linewidth]{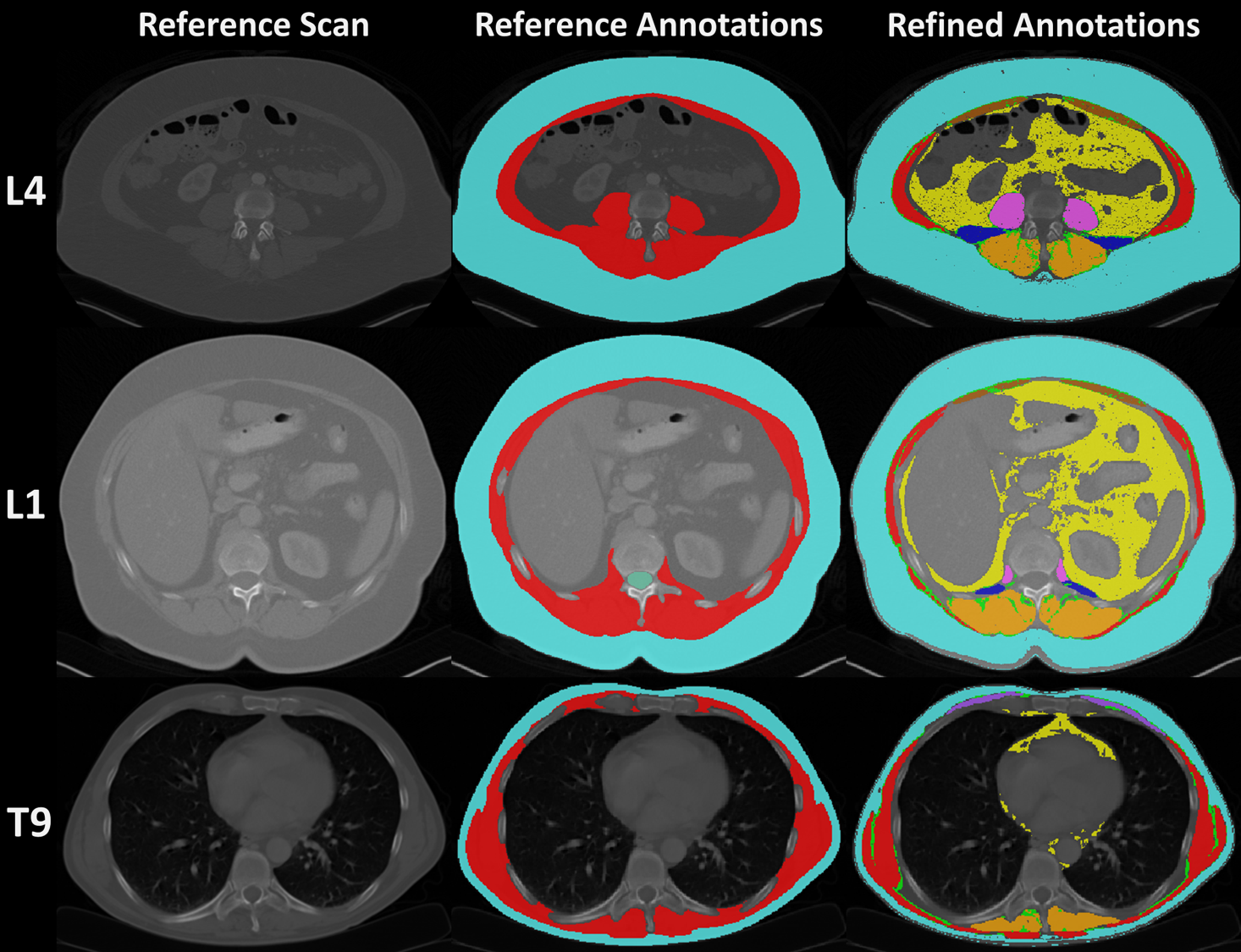}
    \caption{Example axial slices at three vertebral levels (L4--first, L1--second, and T9--third row), showing the reference SAROS CT scan, the reference annotations, and our refined nine-class reference annotations (left, center, and right column).}
    \label{fig:saros_annotations}
\end{figure}

All scans and segmentation maps were standardized to the Right-Anterior-Superior (RAS) anatomical coordinate system. Scans were then cropped along the longitudinal axis to the levels relevant for body composition analysis. Specifically, between the highest detected thoracic vertebra (up to T1) and the lowest detected lumbar vertebra (up to L4), with per-scan boundaries identified from a pre-existing whole-body segmentation map. This yielded a total of 10,920 slices across the 210 scans. The resulting slice-wise prevalence of each tissue class is shown in Fig.~\ref{fig:class_prevalence}. Fig.~\ref{fig:saros_annotations} shows three representative examples comparing the original scans, reference annotations, and our refined reference annotations across varying vertebral levels.

\subsection{Sampling Strategies}
\label{subsec:sampling}

We evaluated three sampling strategies spanning from class-agnostic to class-structured sampling. Specifically, random sampling drew slices uniformly from the training pool and served as the unweighted control. Weighted sampling operated at the slice level, biasing draws toward slices containing rare classes, but without constraining the within-batch composition. Episodic sampling promoted structured class composition within each mini-batch by sampling random subsets of foreground classes across consecutive iterations. The three strategies differed solely in how slices are drawn, with training performed fully supervised and the network architecture, loss function, and optimization settings held constant across all conditions.

\paragraph{Random Sampling.} 
Slices are drawn uniformly from the training set, such that each mini-batch of size $B$ is an i.i.d.\ sample from the full training pool.

\paragraph{Weighted Sampling.} 
Each slice $i$ is assigned a sampling probability, proportional to the inverse frequency of the rarest present foreground class:
\begin{equation}
    p_i \propto \frac{1}{f_{c^*_i}}, \quad c^*_i = \arg\min_{c \in \mathcal{C}_i} f_c,
\end{equation}
where $\mathcal{C}_i$ is the set of classes present in slice $i$ and $f_c$ the frequency of class $c$ across the training set.

\paragraph{Episodic Sampling.}
Each mini-batch is constructed as an episode. In each episode, $N_C$ foreground classes are sampled, and for each class, $N_S$ support slices and $N_Q$ query slices are drawn. Support slices are sampled from the pool of slices containing the target class, while query slices are drawn from the same class-restricted pool. Since classes are sampled uniformly rather than in proportion to their frequency, rare and frequent classes appear as episode targets with equal probability, yielding approximately balanced class exposure over the course of training. The model can then be trained on either the support or the query slices, using the full multi-class labels in both cases.

\subsection{Network Architecture \& Training Protocol}
\label{subsec:architecture}

Inputs were preprocessed by windowing HU (width 400, level 40) followed by linear normalization to $[-1, 1]$. For all experiments, we used a baseline 2D U-Net adopted from the nnU-Net implementation \citep{Isensee2020}. The encoder consisted of six levels with convolutional pooling, beginning at a base feature width of 32 channels and doubling at each subsequent level up to a maximum of 480. Each level contained two convolutional blocks comprising $3 \times 3$ convolutions, instance normalization, and leaky ReLU activation (negative slope $10^{-2}$). The decoder mirrored the encoder, using convolutional upsampling with skip connections and dropout ($p = 0.1$). The output channels corresponded to the nine tissue classes and background. 

We used the AdamW optimizer \citep{loshchilov2019decoupledweightdecayregularization} with an initial learning rate of $10^{-4}$ and weight decay of $10^{-2}$. The learning rate was reduced by a factor of 0.1 at epochs 30 and 45 using a MultiStepLR scheduler. For random and weighted sampling, the batch size was set to 16. For episodic sampling, training comprised 500 episodes per epoch, with each episode sampling $N_C=2$ foreground classes and drawing $N_S=3$ support and $N_Q=3$ query slices per class. Models were trained for a maximum of 200 epochs with early stopping triggered by mean foreground validation Dice (patience of 20 epochs). The loss function combined cross-entropy and Dice loss with equal weighting. 

Segmentation performance was evaluated using two complementary metrics: the Dice similarity coefficient for quantifying area overlap \citep{dice1945measures}, and the 95th-percentile Hausdorff Distance (HD95) for quantifying boundary accuracy \citep{Taha2015}. Metrics were computed per class across all foreground classes.

All experiments were implemented in PyTorch and performed on NVIDIA V100 GPUs with 32\,GB VRAM. The code is available at \url{https://github.com/iasonsky/episodic-sampling}.

\subsection{Experiments}
\label{subsec:exp}

Data were split into 85\% for development and 15\% for testing. The development set comprised 144 scans for training and 36 for validation, with five-fold cross-validation applied at the patient level. The test set comprised 30 held-out scans. To assess whether episodic sampling yields greater benefit under data scarcity and more severe class imbalance, experiments were conducted under two data regimes: (i) a full-data regime (100\%), using all 144 training and 36 validation scans, and (ii) a low-data regime, retaining 10\% of training and validation scans via random subsampling at the patient level. 

As detailed below, in the full-data regime, all sampling strategies required a comparable number of iterations per epoch. As a result, epoch-based scheduling decisions, including learning rate milestones and early stopping patience, corresponded to similar iteration budgets across strategies. In the low-data regime, a 12$\times$ disparity arose between random/weighted and episodic sampling. Learning rate milestones at epoch 30 corresponded to 1,290 iterations under random and weighted sampling versus 15,000 under episodic, and early stopping patience of 20 epochs corresponded to 860 versus 10,000 iterations, respectively.

\begin{itemize}[nosep]
    \item Random/weighted, full-data regime: $\lceil 8{,}369 / 16 \rceil = 523$ iterations per epoch.
    \item Random/weighted, low-data regime: $\lceil 684 / 16 \rceil = 43$ iterations per epoch.
    \item Episodic, both regimes: 500 iterations per epoch (fixed by the number of episodes).
\end{itemize}

Therefore, for a fair comparison in the low-data regime, we disentangled the sampling mechanism from the training budget and systematically evaluated performance under equivalent iterations.

\subsubsection{Fixed iterations with constant learning rate.}
\label{subsubsec:fixed_iters}

We evaluated the per-iteration effectiveness of each sampling strategy by equalizing the training iterations and removing all epoch-based scheduling decisions. To that end, we trained all three samplers for exactly 3{,}000 iterations with a constant learning rate and without early stopping.

\subsubsection{Iteration-calibrated schedule.}
\label{subsubsec:iter_calib}

We tested whether random and weighted sampling can match episodic performance under the same effective training budget. To that end, we rescaled the random and weighted schedules from epoch-based to iteration-equivalent specifications, using episodic's 500 iterations per epoch as a reference. In episodic sampling, milestones at epochs 30 and 45 correspond to 15{,}000 and 22{,}500 iterations, the patience of 20 epochs to 10{,}000 iterations, and the maximum of 200 epochs to 100{,}000 iterations. For random and weighted sampling this corresponded to milestones at epochs 349 and 523, patience of 233 epochs, and a maximum of 2{,}500 epochs.

\section{Results}

\begin{figure}[htp]
    \centering
    \begin{subfigure}[b]{0.8\textwidth}
        \centering
        \includegraphics[width=\textwidth]{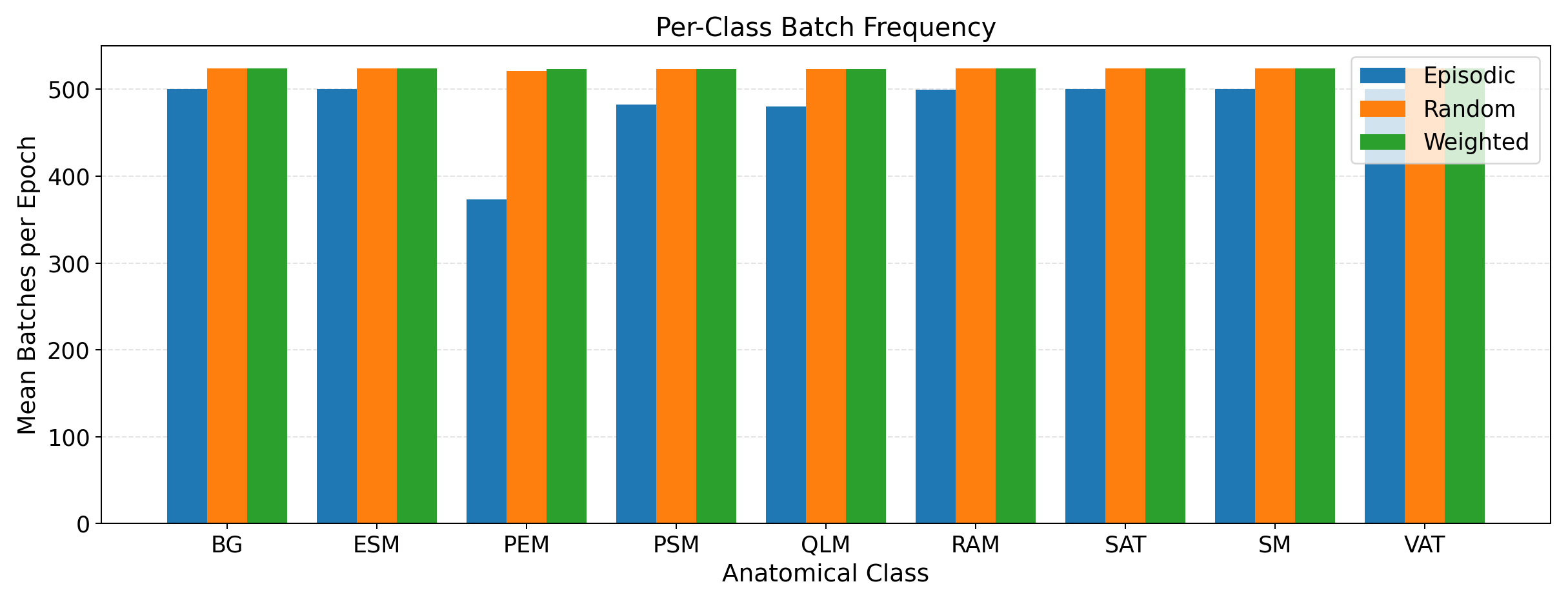}
        \caption{Full-Data (100\%) Training}
        \label{fig:sampling_100}
    \end{subfigure}
    \vspace{0.5cm}
    \begin{subfigure}[b]{0.8\textwidth}
        \centering
        \includegraphics[width=\textwidth]{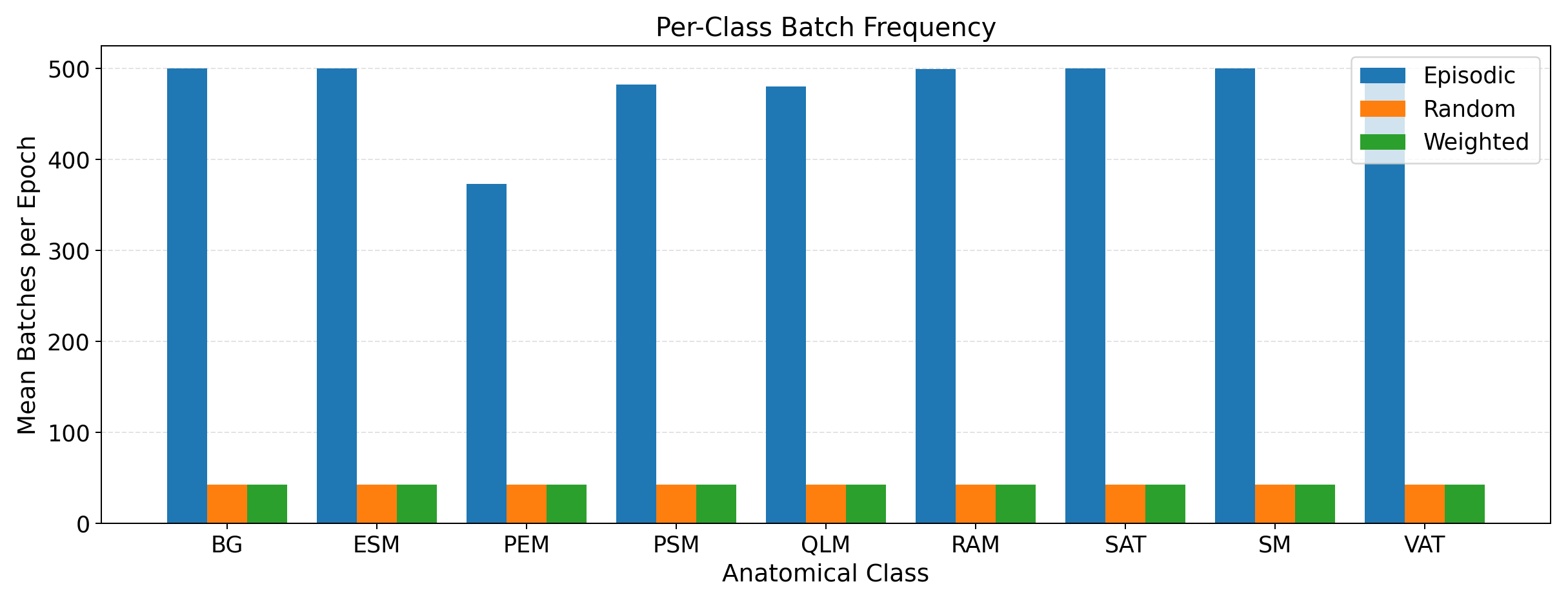}
        \caption{Low-Data (10\%) Training}
        \label{fig:samling_10}
    \end{subfigure}
    \caption{Per-class batch frequency per epoch for episodic (blue), random (orange), and weighted (green) sampling for a representative fold (fold~0) under full-data (a) and low-data (b). Random and weighted sampling scale down proportionally with reduced dataset size. Episodic sampling maintains ${\sim}500$ batches per class per epoch in both regimes.}
    \label{fig:class_sampling_distribution}
\end{figure}

\begin{figure}[htp]
    \centering
    \begin{subfigure}[b]{\textwidth}
        \centering
        \includegraphics[width=\textwidth]{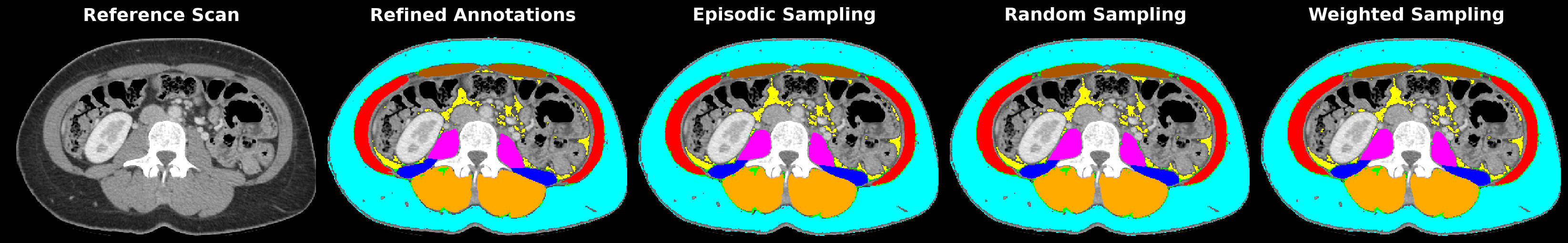}
        \caption{Full-Data (100\%) Training}
        \label{fig:qualitative_full}
    \end{subfigure}
    \vspace{0.3em}
    \begin{subfigure}[b]{\textwidth}
        \centering
        \includegraphics[width=\textwidth]{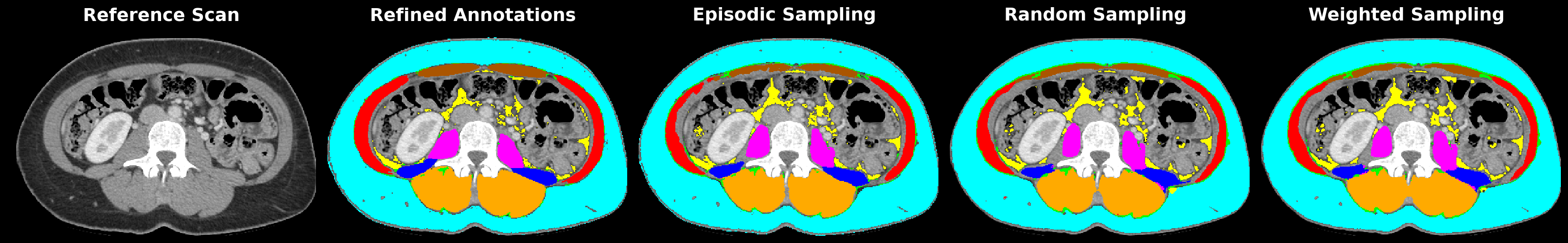}
        \caption{Low-Data (10\%) Training}
        \label{fig:qualitative_low}
    \end{subfigure}
    \caption{Representative segmentations at the L3 vertebral level for a test case under standard epoch-based full-data (a) and low-data (b) training. Each column (left to right) shows the reference CT scan, refined annotations, and predictions from episodic, random, and weighted sampling, respectively.}
    \label{fig:qualitative_epoch_based_training}
\end{figure}

Table~\ref{tab:main_comparison} compares the performance of episodic, random, and weighted sampling under both the full-data (100\%) and low-data (10\%) regimes. In the full-data regime, the choice of sampling strategy had minimal impact. Episodic achieved a mean Dice of $0.882$, compared to $0.878$ for both random and weighted, with a corresponding advantage in HD95 ($6.77$~mm vs.\ $7.98$~mm and $7.80$~mm). This modest effect was consistent with the near-matched iteration budgets across strategies in this regime (523 vs.\ 500 iterations per epoch). In the low-data regime, the advantage of episodic became pronounced, with a mean Dice of $0.787$, compared to $0.758$ for random and $0.762$ for weighted sampling. Performance improved on eight of the nine foreground classes, with the largest gains observed on the least prevalent classes (IMAT, QLM, PEM, PSM; Fig.~\ref{fig:class_prevalence}). In addition, episodic achieved the lowest HD95 on seven of nine classes, while random sampling achieved the best average HD95 ($15.89$~mm vs.\ $16.05$~mm for episodic). However, as detailed in Sec.~\ref{subsec:exp} and shown in Fig.~\ref{fig:class_sampling_distribution}, episodic sampling ran 12$\times$ more training iterations per epoch in the low-data regime. Across both regimes, random and weighted sampling performed comparably, with neither consistently outperforming the other. Fig.~\ref{fig:qualitative_epoch_based_training} shows representative segmentations for each sampling strategy under both training regimes.

For episodic sampling, we ran an ablation study comparing using the supports and queries as training inputs. As shown in the Appendix Table~\ref{app:appendix}, query- or support-based supervision yielded near-identical performance in both data regimes, with a slight advantage on queries (mean Dice of $0.882$ [queries] vs.\ $0.881$ [supports] at 100\%, both $0.787$ at 10\%). Therefore, for the rest of our experiments, we selected query-based supervision.

\begin{table}[htp]
\centering
\caption{Performance of episodic, random, and weighted sampling on the held-out test set under full-data (100\%) and low-data (10\%) training. Best scores are highlighted in bold. 
}
\label{tab:main_comparison}
\resizebox{\textwidth}{!}{%
\begin{tabular}{l ccc ccc}
\toprule
& \multicolumn{3}{c}{Dice Scores $\uparrow$} & \multicolumn{3}{c}{HD95 $\downarrow$} \\
\cmidrule(lr){2-4} \cmidrule(lr){5-7}
Class & Episodic & Random & Weighted & Episodic & Random & Weighted \\
\midrule
& \multicolumn{6}{c}{\textit{Full-Data (100\%) Training}} \\
\midrule
AVERAGE & $\mathbf{0.882} \pm 0.13$ & $0.878 \pm 0.14$ & $0.878 \pm 0.14$ & $\mathbf{6.77} \pm 14.07$ & $7.98 \pm 22.75$ & $7.80 \pm 18.77$ \\
\addlinespace
ESM & $\mathbf{0.967} \pm 0.02$ & $\mathbf{0.967} \pm 0.02$ & $\mathbf{0.967} \pm 0.02$ & $\mathbf{0.49} \pm 0.90$ & $0.52 \pm 0.93$ & $0.53 \pm 0.92$ \\
IMAT & $\mathbf{0.783} \pm 0.06$ & $0.782 \pm 0.06$ & $0.781 \pm 0.06$ & $\mathbf{8.74} \pm 5.58$ & $8.81 \pm 5.60$ & $8.79 \pm 5.47$ \\
PEM & $\mathbf{0.646} \pm 0.23$ & $0.621 \pm 0.25$ & $0.620 \pm 0.25$ & $\mathbf{32.56} \pm 30.90$ & $43.87 \pm 57.56$ & $41.16 \pm 43.69$ \\
PSM & $\mathbf{0.903} \pm 0.05$ & $0.901 \pm 0.05$ & $\mathbf{0.903} \pm 0.05$ & $\mathbf{4.67} \pm 4.44$ & $5.08 \pm 4.78$ & $5.18 \pm 4.90$ \\
QLM & $\mathbf{0.860} \pm 0.06$ & $0.855 \pm 0.06$ & $0.853 \pm 0.06$ & $6.90 \pm 5.80$ & $\mathbf{6.86} \pm 6.27$ & $7.31 \pm 7.09$ \\
RAM & $\mathbf{0.904} \pm 0.03$ & $0.902 \pm 0.04$ & $0.901 \pm 0.04$ & $4.87 \pm 3.79$ & $\mathbf{4.49} \pm 3.28$ & $5.01 \pm 3.47$ \\
SAT & $\mathbf{0.983} \pm 0.01$ & $\mathbf{0.983} \pm 0.01$ & $\mathbf{0.983} \pm 0.01$ & $0.22 \pm 0.91$ & $\mathbf{0.20} \pm 0.91$ & $\mathbf{0.20} \pm 0.92$ \\
SM & $\mathbf{0.936} \pm 0.02$ & $\mathbf{0.936} \pm 0.02$ & $0.935 \pm 0.02$ & $\mathbf{2.85} \pm 3.84$ & $2.98 \pm 3.73$ & $2.89 \pm 3.62$ \\
VAT & $0.948 \pm 0.02$ & $\mathbf{0.950} \pm 0.03$ & $0.949 \pm 0.03$ & $\mathbf{1.34} \pm 1.58$ & $1.37 \pm 1.69$ & $\mathbf{1.34} \pm 1.74$ \\
\midrule
& \multicolumn{6}{c}{\textit{Low-Data (10\%) Training}} \\
\midrule
AVERAGE & $\mathbf{0.787} \pm 0.17$ & $0.758 \pm 0.18$ & $0.762 \pm 0.18$ & $16.05 \pm 27.56$ & $\mathbf{15.89} \pm 20.52$ & $17.02 \pm 26.35$ \\
\addlinespace
ESM & $\mathbf{0.936} \pm 0.03$ & $0.923 \pm 0.03$ & $0.923 \pm 0.03$ & $3.08 \pm 4.91$ & $\mathbf{2.91} \pm 2.57$ & $3.04 \pm 2.49$ \\
IMAT & $\mathbf{0.667} \pm 0.11$ & $0.618 \pm 0.11$ & $0.617 \pm 0.10$ & $\mathbf{12.99} \pm 9.47$ & $14.55 \pm 11.27$ & $14.24 \pm 10.43$ \\
PEM & $\mathbf{0.519} \pm 0.22$ & $0.483 \pm 0.20$ & $0.495 \pm 0.21$ & $62.10 \pm 58.44$ & $\mathbf{50.88} \pm 34.00$ & $62.43 \pm 52.99$ \\
PSM & $\mathbf{0.805} \pm 0.10$ & $0.756 \pm 0.12$ & $0.773 \pm 0.11$ & $\mathbf{11.93} \pm 8.95$ & $14.07 \pm 9.96$ & $14.05 \pm 9.15$ \\
QLM & $\mathbf{0.670} \pm 0.13$ & $0.632 \pm 0.14$ & $0.646 \pm 0.13$ & $\mathbf{18.75} \pm 13.57$ & $18.96 \pm 11.48$ & $20.09 \pm 11.97$ \\
RAM & $\mathbf{0.772} \pm 0.07$ & $0.756 \pm 0.07$ & $0.751 \pm 0.07$ & $\mathbf{14.14} \pm 7.30$ & $17.48 \pm 7.25$ & $15.72 \pm 6.86$ \\
SAT & $\mathbf{0.969} \pm 0.02$ & $0.956 \pm 0.04$ & $0.955 \pm 0.04$ & $\mathbf{0.82} \pm 1.47$ & $0.84 \pm 1.47$ & $0.94 \pm 1.78$ \\
SM & $\mathbf{0.849} \pm 0.07$ & $0.833 \pm 0.08$ & $0.834 \pm 0.08$ & $\mathbf{11.97} \pm 18.39$ & $13.42 \pm 18.74$ & $13.23 \pm 18.53$ \\
VAT & $\mathbf{0.890} \pm 0.07$ & $0.860 \pm 0.08$ & $0.855 \pm 0.09$ & $\mathbf{10.17} \pm 14.06$ & $11.08 \pm 16.23$ & $10.94 \pm 15.85$ \\
\bottomrule
\end{tabular}
}
\end{table}

\begin{figure}[htp]
    \centering
    \begin{subfigure}[b]{\textwidth}
        \centering
        \includegraphics[width=\textwidth]{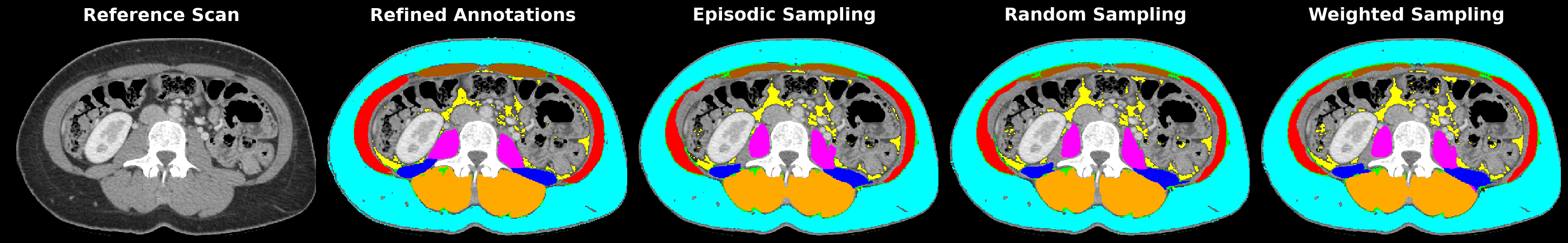}
        \caption{Low-Data (10\%) Fixed-Iteration Training}
        \label{fig:qualitative_fixed}
    \end{subfigure}
    \vspace{0.3em}
    \begin{subfigure}[b]{\textwidth}
        \centering
        \includegraphics[width=\textwidth]{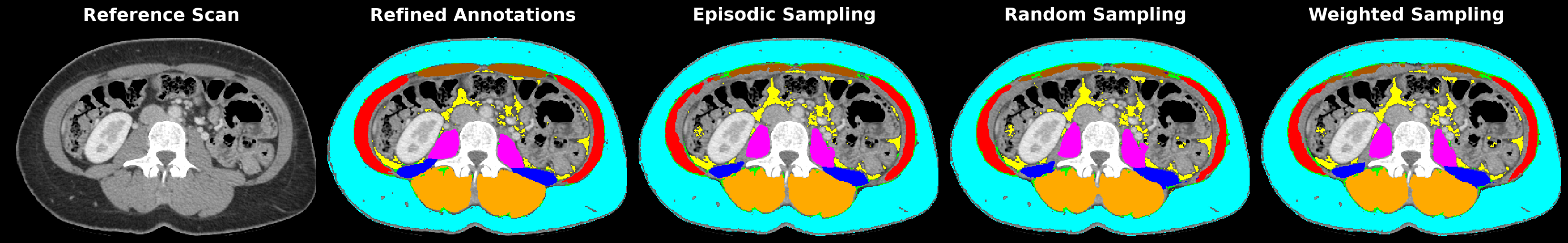}
        \caption{Low-Data (10\%) Iteration-Calibrated Training}
        \label{fig:qualitative_calibrated}
    \end{subfigure}
    \caption{Representative segmentations at the L3 vertebral level for a test case under low-data fixed-iteration (a) and iteration-calibrated (b) training. Each column (left to right) shows the reference CT scan, refined annotations, and predictions from episodic, random, and weighted sampling, respectively.}
\end{figure}

Table~\ref{tab:controlled_comparison} reports the results of the fixed 3{,}000 iterations with constant learning rate experiment (Sec.~\ref{subsubsec:fixed_iters}). All three samplers converged to similar performance ($0.778$ vs.\ $0.773$ vs.\ $0.773$ mean Dice for episodic, random, and weighted, respectively). Episodic sampling achieved the highest Dice on five of nine classes (ESM, IMAT, PSM, RAM, SM), while weighted led on PEM and QLM, and random on SAT and VAT. For HD95, weighted achieved the best average (15.09~mm), followed by random (15.55~mm) and episodic (15.70~mm), with no single sampler consistently dominating across classes. Fig.~\ref{fig:qualitative_fixed} shows representative segmentations for each sampling strategy under low-data fixed-iteration training.

\begin{table}[htbp]
\centering
\caption{Performance of episodic, random, and weighted sampling on the held-out test set under low-data (10\%) fixed-iteration training (Sec.~\ref{subsubsec:fixed_iters}). Best scores are highlighted in bold.}
\label{tab:controlled_comparison}
\resizebox{\textwidth}{!}{%
\begin{tabular}{l ccc ccc}
\toprule
& \multicolumn{3}{c}{Dice Scores $\uparrow$} & \multicolumn{3}{c}{Hausdorff Scores (95th) $\downarrow$} \\
\cmidrule(lr){2-4} \cmidrule(lr){5-7}
Class & Episodic & Random & Weighted & Episodic & Random & Weighted \\
\midrule
& \multicolumn{6}{c}{\textit{Low-Data (10\%) Training}} \\
\midrule
AVERAGE & $\mathbf{0.778} \pm 0.17$ & $0.773 \pm 0.17$ & $0.773 \pm 0.17$ & $15.70 \pm 26.50$ & $15.55 \pm 20.80$ & $\mathbf{15.09} \pm 19.12$ \\
\addlinespace
ESM & $\mathbf{0.932} \pm 0.03$ & $0.926 \pm 0.03$ & $0.927 \pm 0.03$ & $3.12 \pm 5.05$ & $3.66 \pm 4.60$ & $\mathbf{3.04} \pm 2.76$ \\
IMAT & $\mathbf{0.657} \pm 0.10$ & $0.650 \pm 0.10$ & $0.637 \pm 0.11$ & $\mathbf{13.03} \pm 9.20$ & $13.29 \pm 8.70$ & $13.28 \pm 9.02$ \\
PEM & $0.514 \pm 0.22$ & $0.517 \pm 0.22$ & $\mathbf{0.523} \pm 0.21$ & $56.21 \pm 57.91$ & $52.40 \pm 35.53$ & $\mathbf{48.95} \pm 30.63$ \\
PSM & $\mathbf{0.783} \pm 0.11$ & $0.773 \pm 0.11$ & $0.782 \pm 0.10$ & $\mathbf{12.80} \pm 9.43$ & $13.13 \pm 8.34$ & $13.35 \pm 8.99$ \\
QLM & $0.648 \pm 0.14$ & $0.650 \pm 0.13$ & $\mathbf{0.657} \pm 0.13$ & $20.35 \pm 14.28$ & $20.10 \pm 11.66$ & $\mathbf{19.29} \pm 11.27$ \\
RAM & $\mathbf{0.777} \pm 0.07$ & $0.754 \pm 0.07$ & $0.757 \pm 0.07$ & $\mathbf{14.20} \pm 7.05$ & $15.45 \pm 7.90$ & $15.85 \pm 7.36$ \\
SAT & $0.960 \pm 0.03$ & $\mathbf{0.963} \pm 0.03$ & $0.960 \pm 0.03$ & $0.85 \pm 1.54$ & $\mathbf{0.73} \pm 1.34$ & $0.84 \pm 1.56$ \\
SM & $\mathbf{0.845} \pm 0.07$ & $0.838 \pm 0.07$ & $0.833 \pm 0.08$ & $12.91 \pm 18.49$ & $\mathbf{12.71} \pm 18.45$ & $13.08 \pm 18.55$ \\
VAT & $0.875 \pm 0.07$ & $\mathbf{0.878} \pm 0.07$ & $0.871 \pm 0.08$ & $\mathbf{9.15} \pm 14.30$ & $9.70 \pm 14.34$ & $10.34 \pm 14.92$ \\
\bottomrule
\end{tabular}
}
\end{table}

Table~\ref{tab:calibrated_comparison} reports the results of the iteration-calibrated schedule experiment (Sec.~\ref{subsubsec:iter_calib}). Iteration calibration substantially closed the gap between samplers. Random sampling improved from $0.758$ mean Dice (Table~\ref{tab:main_comparison}) to $0.777$, and weighted from $0.762$ to $0.778$. The largest margins between episodic and random samplings appeared on RAM ($+3.0$~pp), QLM ($+2.0$~pp), and PSM ($+1.3$~pp). The iteration-calibrated schedule also narrowed the HD95 gap, with episodic achieving the lowest HD95 on seven of nine classes, yet random still led on average ($15.95$~mm vs.\ $16.05$~mm for episodic). This was attributed largely to PEM performance ($62.10$~mm for episodic vs.\ $55.77$~mm for random). Fig.~\ref{fig:qualitative_calibrated} shows representative segmentations for each sampling strategy under low-data iteration-calibrated training.

\begin{table}[htp]
\centering
\caption{Performance of episodic, random, and weighted sampling on the held-out test set under low-data (10\%) iteration-calibrated training (Sec.~\ref{subsubsec:iter_calib}). Best scores are highlighted in bold.}
\resizebox{\textwidth}{!}{%
\label{tab:calibrated_comparison}
\begin{tabular}{l ccc ccc}
\toprule
& \multicolumn{3}{c}{Dice Scores $\uparrow$} & \multicolumn{3}{c}{Hausdorff Scores (95th) $\downarrow$} \\
\cmidrule(lr){2-4} \cmidrule(lr){5-7}
Class & Episodic & Random & Weighted & Episodic & Random & Weighted \\
\midrule
& \multicolumn{6}{c}{\textit{Low-Data (10\%) Training}} \\
\midrule
AVERAGE & $\mathbf{0.787} \pm 0.17$ & $0.777 \pm 0.17$ & $0.778 \pm 0.17$ & $16.05 \pm 27.56$ & $\mathbf{15.95} \pm 24.38$ & $16.23 \pm 23.64$ \\
\addlinespace
ESM & $\mathbf{0.936} \pm 0.03$ & $0.929 \pm 0.03$ & $0.930 \pm 0.03$ & $\mathbf{3.08} \pm 4.91$ & $3.48 \pm 4.16$ & $4.37 \pm 5.42$ \\
IMAT & $\mathbf{0.667} \pm 0.11$ & $0.659 \pm 0.10$ & $0.658 \pm 0.11$ & $\mathbf{12.99} \pm 9.47$ & $13.18 \pm 8.59$ & $13.28 \pm 9.19$ \\
PEM & $0.519 \pm 0.22$ & $\mathbf{0.521} \pm 0.20$ & $0.516 \pm 0.21$ & $62.10 \pm 58.44$ & $\mathbf{55.77} \pm 49.25$ & $57.23 \pm 44.97$ \\
PSM & $\mathbf{0.805} \pm 0.10$ & $0.792 \pm 0.10$ & $0.795 \pm 0.10$ & $\mathbf{11.93} \pm 8.95$ & $12.24 \pm 8.78$ & $12.06 \pm 7.93$ \\
QLM & $\mathbf{0.670} \pm 0.13$ & $0.650 \pm 0.14$ & $0.661 \pm 0.13$ & $\mathbf{18.75} \pm 13.57$ & $19.84 \pm 12.95$ & $19.39 \pm 12.58$ \\
RAM & $\mathbf{0.772} \pm 0.07$ & $0.742 \pm 0.08$ & $0.743 \pm 0.08$ & $\mathbf{14.14} \pm 7.30$ & $16.38 \pm 8.66$ & $16.19 \pm 8.00$ \\
SAT & $\mathbf{0.969} \pm 0.02$ & $0.968 \pm 0.03$ & $0.968 \pm 0.03$ & $0.82 \pm 1.47$ & $\mathbf{0.81} \pm 1.50$ & $0.86 \pm 1.59$ \\
SM & $\mathbf{0.849} \pm 0.07$ & $0.840 \pm 0.08$ & $0.839 \pm 0.08$ & $\mathbf{11.97} \pm 18.39$ & $12.65 \pm 18.52$ & $12.70 \pm 18.42$ \\
VAT & $\mathbf{0.890} \pm 0.07$ & $0.882 \pm 0.08$ & $0.882 \pm 0.08$ & $\mathbf{10.17} \pm 14.06$ & $10.53 \pm 15.24$ & $11.34 \pm 15.50$ \\
\bottomrule
\end{tabular}
}
\end{table}

Training dynamics across all protocols were consistent with these results (Appendix~\ref{app:appendix}). In the full-data regime (Fig.~\ref{fig:experiment_100_percent}), all three samplers converged comparably, consistent with their near-matched iteration budgets (Sec.~\ref{subsec:exp}). In the uncalibrated low-data regime (Fig.~\ref{fig:experiment_10_uncalibrated}), random and weighted sampling terminated at ${\sim}2{,}500$ iterations via early stopping while episodic ran to ${\sim}30{,}000$. Under the fixed $3{,}000$-iteration protocol (Fig.~\ref{fig:experiment_10_fixed}), all three samplers reached similar Dice but episodic's rare-class curves climbed later, indicating incomplete class coverage at this budget. Under the iteration-calibrated schedule (Fig.~\ref{fig:experiment_10_calibrated}), random and weighted peaked at $5{,}100$ and $7{,}800$ iterations and overfit thereafter, while episodic peaked at ${\sim}15{,}000$ with validation loss remaining below the other two throughout the budget.

\section{Discussion}

In this work, we evaluated episodic sampling as a model-agnostic, plug-and-play batch construction strategy for class-imbalanced muscle and adipose tissue segmentation in CT. We decoupled episodic batch construction from metric-based learning \citep{snell2017prototypicalnetworksfewshotlearning,Ouyang2020} and compared it against random and weighted sampling, under both full- and low-data training regimes (Sec.~\ref{subsec:exp}). Our work establishes episodic sampling as a feasible supervised-training mechanism, while exposing the intrinsic relationships between number of epochs, training iterations, learning rate milestones, and early stopping patience as an under-recognized confound.

We showed that sampling strategies are heavily influenced by this confound in epoch-based training. When iteration counts per epoch differ, epoch-tied scheduling decisions translate into substantially different effective training budgets. Once the confound is controlled, the apparent benefit of class-aware sampling diminishes and performance gains are mostly attributable to increased iteration budget. Under a matched fixed-iteration budget, the three samplers  performed comparably (Table~\ref{tab:controlled_comparison}), and  even under the iteration-calibrated schedule, episodic sampling's advantage narrowed from 2.9 to 1.0 percentage points (Table~\ref{tab:calibrated_comparison}). The residual advantage 
coincided with delayed overfitting: random and weighted sampling reached their best checkpoints earlier and showed rising validation loss thereafter, while episodic sampling continued improving for approximately three times as many iterations. This pattern is 
consistent with an implicit regularization effect of class-balanced batch construction, though the mechanism remains unclear without direct measurements of gradient variance or feature-space geometry.

Notably, weighted sampling performed comparably to random sampling across all conditions, with neither consistently outperforming the other. Both strategies adjust slice-level sampling probabilities but leave the pixel-level class composition within each selected slice unchanged. This suggests that slice-level frequency rebalancing alone is insufficient when the dominant source of imbalance is within image rather than across-image.

Yet these observations raise concerns about epoch-based training and comparisons, which usually attribute gains to the sampling mechanism or the network architecture \citep{arazo2021importantimportancesamplingdeep,shwartzziv2023simplifyingneuralnetworktraining}. The state-of-the-art nnU-Net framework circumvents this confound by fixing the training iterations to a high number ($250{,}000$) regardless of dataset size \citep{Isensee2020}. Here, we identify that the confound is not specific to the sampling mechanism, but applies to any epoch-based scheduler. Therefore, our work recommends scaling the training schedule to the dataset size by adjusting the iterations of the schedule. In practice, this means disentangling the sampling mechanism from the training budget, a step that can also enhance reproducibility \citep{li2020budgetedtrainingrethinkingdeep}.

Episodic sampling has direct implications for body composition analysis, where accurate segmentation of small muscle subgroups, like psoas and quadratus lumborum, and intermuscular adipose tissue is highly challenging. Body composition pipelines typically operate on a single 2D slice or collapse the problem to coarse tissue labels \citep{blankemeier_comp2comp_2023,haubold2024boa}, leaving rare classes unaddressed. Thus, episodic sampling offers a low-cost, model-agnostic improvement directly at the input level, without modifying the loss or requiring additional annotations. 

Finally, three methodological choices are worth noting regarding their impact on our results. Reference annotations were refined using automated body segmentation models \citep{haubold2024boa}, which may introduce label noise warranting multi-seed cross-validation \citep{bouthillier2021accountingvariancemachinelearning}. The calibrated schedule was anchored to the 500 iterations per epoch of episodic sampling, without exploring alternative reference budgets. Additionally, our evaluation was limited to a single task, baseline model, and loss configuration, though since the budget decomposition is mechanism-driven rather than task-specific, the findings are likely to transfer to other class-imbalanced settings. Future work could assess why episodic sampling remains advantageous at matched iteration budgets, through gradient-variance analysis \citep{zhao2014stochastic,DBLP:journals/corr/abs-1803-00942}, feature-space geometry \citep{you2023rethinkingsemisupervisedmedicalimage}, or calibration metrics. Furthermore, combining episodic sampling with random or weighted sampling and loss-based rebalancing strategies could yield complementary gains.

\section{Conclusion}

We evaluated episodic sampling as a plug-and-play strategy for class imbalance in medical image segmentation, comparing it against standard random and weighted sampling. We showed that the apparent advantage of episodic sampling stemmed primarily from epoch-based scheduling, rather than the sampling mechanism itself. Even at matched training budgets, episodic sampling retained a small advantage, with random and weighted sampling overfitting earlier under extended training. Our findings exposed a confound in sampling-strategy comparisons, motivating iteration-aware protocols on small datasets to distinguish true algorithmic improvements from incidental compute differences. Future work could explore curriculum frameworks that combine episodic sampling with random or weighted sampling, or with loss-based rebalancing strategies, and develop systematic heuristics for dataset-adaptive iteration budgets.

\section*{Acknowledgments}

This work is supported by the Artillery Consortium and European Union’s Horizon Europe under Grant No. 101080983. 

\bibliographystyle{unsrtnat}
\bibliography{references}

\newpage
\appendix
\section{Appendix}
\label{app:appendix}

\begin{table}[htpt]
\centering
\caption{Performance of episodic query-based and support-based training on the held-out test set under full-data (100\%) and low-data (10\%). Best scores are highlighted in bold.}
\resizebox{0.8\textwidth}{!}{%
\label{tab:episodic_ablation}
\begin{tabular}{l cc cc}
\toprule
& \multicolumn{2}{c}{Dice Scores $\uparrow$} & \multicolumn{2}{c}{Hausdorff Scores (95th) $\downarrow$} \\
\cmidrule(lr){2-3} \cmidrule(lr){4-5}
Class & Queries & Supports & Queries & Supports \\
\midrule
& \multicolumn{4}{c}{\textit{Full-Data (100\%) Training}} \\
\midrule
AVERAGE & \textbf{0.882} $\pm$ 0.13 & 0.881 $\pm$ 0.13 & \textbf{6.77} $\pm$ 14.07 & 8.28 $\pm$ 20.79 \\
ESM & \textbf{0.967} $\pm$ 0.02 & \textbf{0.967} $\pm$ 0.02 & \textbf{0.49} $\pm$ 0.90 & \textbf{0.49} $\pm$ 0.91 \\
IMAT & \textbf{0.783} $\pm$ 0.06 & 0.782 $\pm$ 0.06 & \textbf{8.74} $\pm$ 5.58 & 8.98 $\pm$ 5.87 \\
PEM & \textbf{0.646} $\pm$ 0.23 & 0.638 $\pm$ 0.24 & \textbf{32.56} $\pm$ 30.90 & 44.76 $\pm$ 48.02 \\
PSM & \textbf{0.903} $\pm$ 0.05 & \textbf{0.903} $\pm$ 0.05 & \textbf{4.67} $\pm$ 4.44 & 4.73 $\pm$ 4.57 \\
QLM & \textbf{0.860} $\pm$ 0.06 & 0.859 $\pm$ 0.06 & \textbf{6.90} $\pm$ 5.80 & 7.28 $\pm$ 6.51 \\
RAM & \textbf{0.904} $\pm$ 0.03 & \textbf{0.904} $\pm$ 0.04 & \textbf{4.87} $\pm$ 3.79 & 4.94 $\pm$ 3.84 \\
SAT & \textbf{0.983} $\pm$ 0.01 & \textbf{0.983} $\pm$ 0.01 & \textbf{0.22} $\pm$ 0.91 & 0.23 $\pm$ 0.91 \\
SM & \textbf{0.936} $\pm$ 0.02 & \textbf{0.936} $\pm$ 0.02 & \textbf{2.85} $\pm$ 3.84 & 2.93 $\pm$ 4.00 \\
VAT & 0.948 $\pm$ 0.02 & \textbf{0.949} $\pm$ 0.02 & \textbf{1.34} $\pm$ 1.58 & 1.39 $\pm$ 1.59 \\
\midrule
& \multicolumn{4}{c}{\textit{Low-Data (10\%) Training}} \\
\midrule
AVERAGE & $\mathbf{0.787} \pm 0.17$ & $\mathbf{0.787} \pm 0.17$ & $16.05 \pm 27.56$ & $\mathbf{15.08} \pm 23.92$ \\
\addlinespace
ESM & $\mathbf{0.936} \pm 0.03$ & $0.933 \pm 0.03$ & $\mathbf{3.08} \pm 4.91$ & $3.16 \pm 3.68$ \\
IMAT & $\mathbf{0.667} \pm 0.11$ & $0.664 \pm 0.11$ & $12.99 \pm 9.47$ & $\mathbf{12.87} \pm 8.91$ \\
PEM & $0.519 \pm 0.22$ & $\mathbf{0.522} \pm 0.21$ & $62.10 \pm 58.44$ & $\mathbf{54.74} \pm 49.65$ \\
PSM & $0.805 \pm 0.10$ & $\mathbf{0.811} \pm 0.09$ & $11.93 \pm 8.95$ & $\mathbf{11.44} \pm 8.26$ \\
QLM & $0.670 \pm 0.13$ & $\mathbf{0.675} \pm 0.13$ & $\mathbf{18.75} \pm 13.57$ & $18.93 \pm 13.37$ \\
RAM & $\mathbf{0.772} \pm 0.07$ & $0.767 \pm 0.07$ & $\mathbf{14.14} \pm 7.30$ & $14.76 \pm 7.54$ \\
SAT & $\mathbf{0.969} \pm 0.02$ & $0.968 \pm 0.03$ & $\mathbf{0.82} \pm 1.47$ & $0.84 \pm 1.57$ \\
SM & $0.849 \pm 0.07$ & $\mathbf{0.850} \pm 0.08$ & $\mathbf{11.97} \pm 18.39$ & $12.30 \pm 18.58$ \\
VAT & $\mathbf{0.890} \pm 0.07$ & $0.888 \pm 0.07$ & $10.17 \pm 14.06$ & $\mathbf{9.32} \pm 13.40$ \\
\bottomrule
\end{tabular}
}
\end{table}

\begin{figure}[htp]
    \centering
    \includegraphics[width=\textwidth]{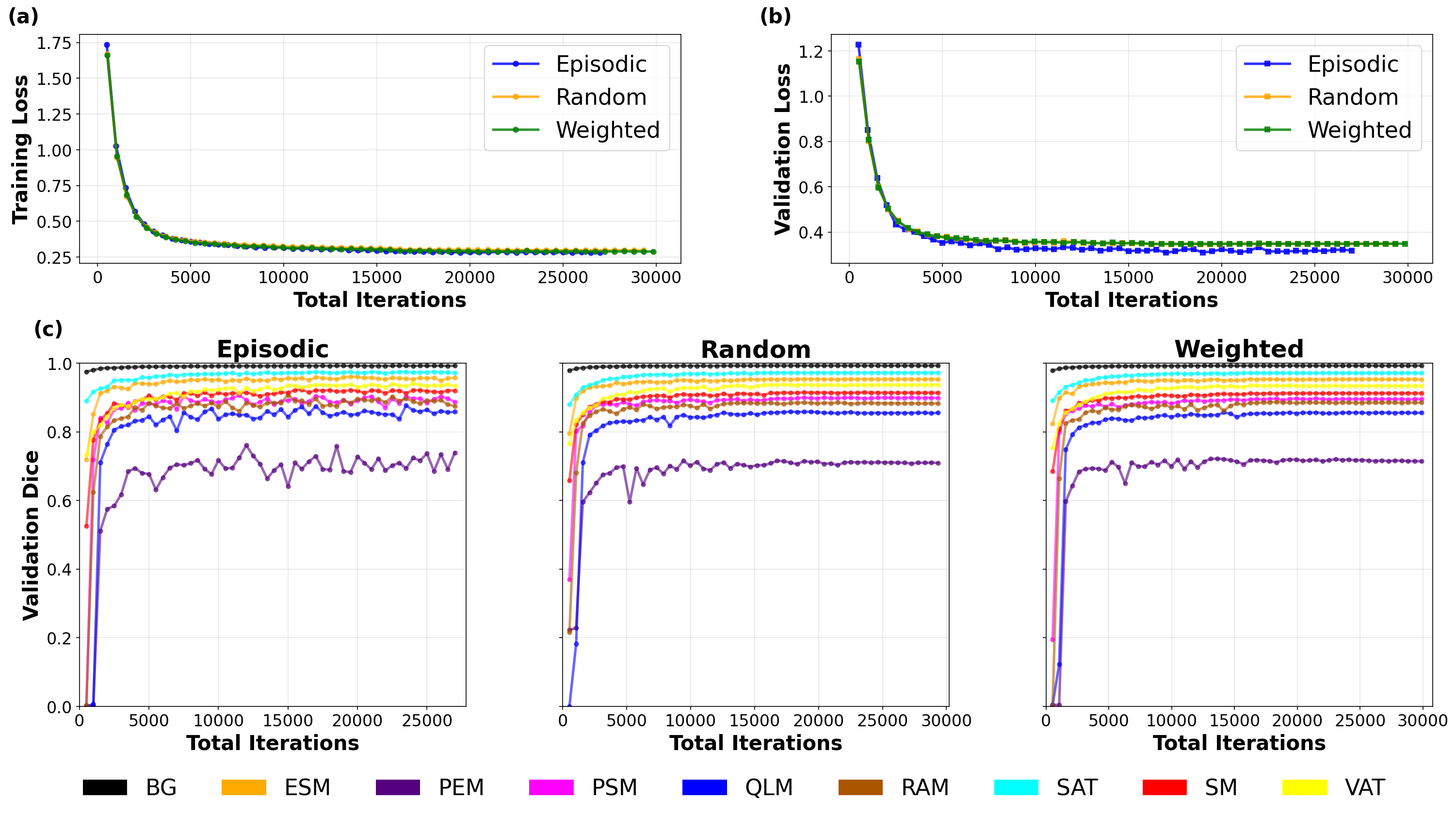}
    \caption{Full-data 100\% regime. (a; top--left) training Loss, (b; top--right) validation Loss, (c; bottom) per-class validation Dice performance for episodic (left), random (middle), and weighted (right).}
    \label{fig:experiment_100_percent}
\end{figure}

\begin{figure}[htp]
    \centering
    \includegraphics[width=\textwidth]{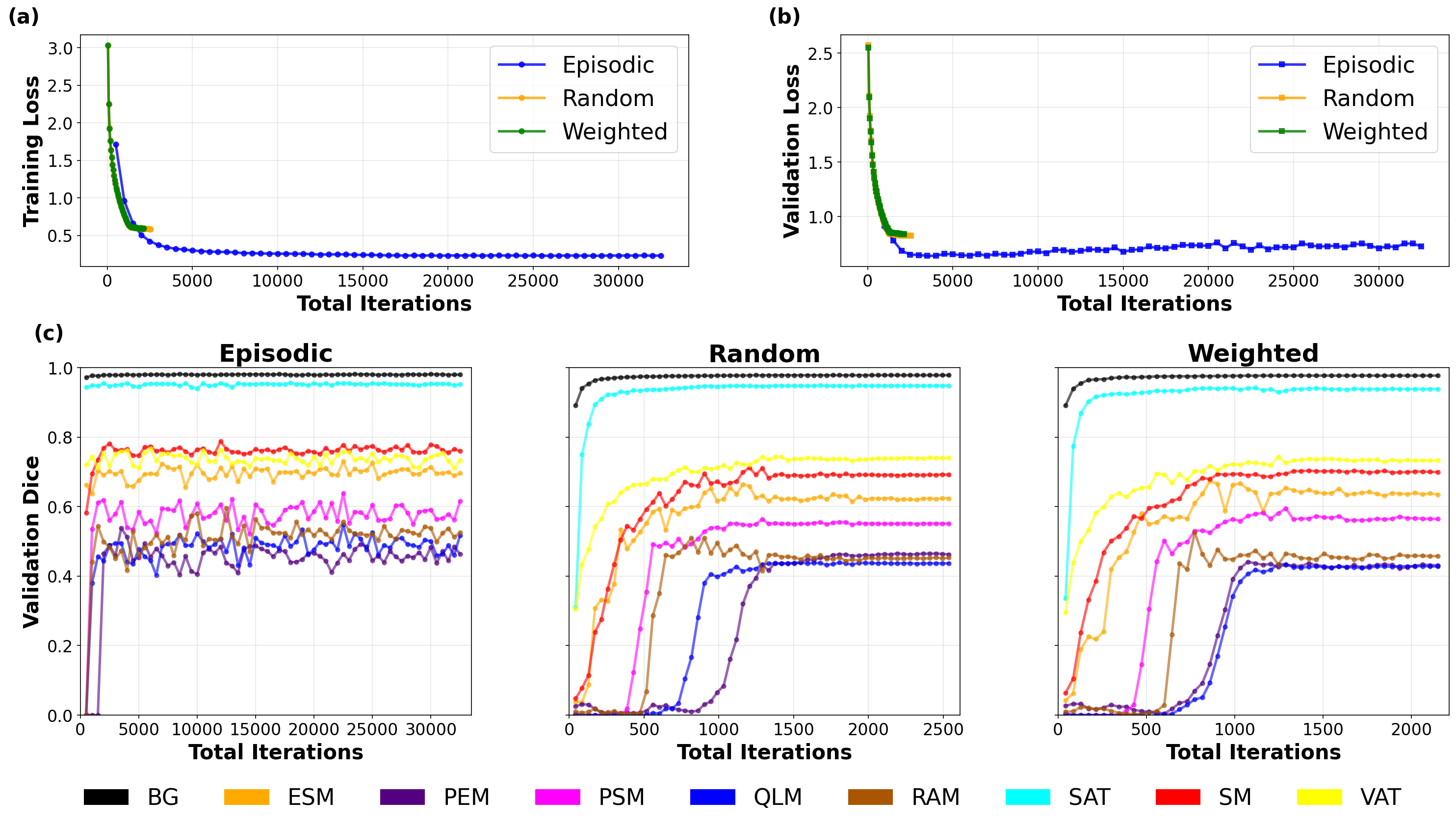}
    \caption{Low-data 10\% regime. (a; top--left) training Loss, (b; top--right) validation Loss, (c; bottom) per-class validation Dice performance for episodic (left), random (middle), and weighted (right).}
    \label{fig:experiment_10_uncalibrated}
\end{figure}

\begin{figure}[htp]
    \centering
    \includegraphics[width=\textwidth]{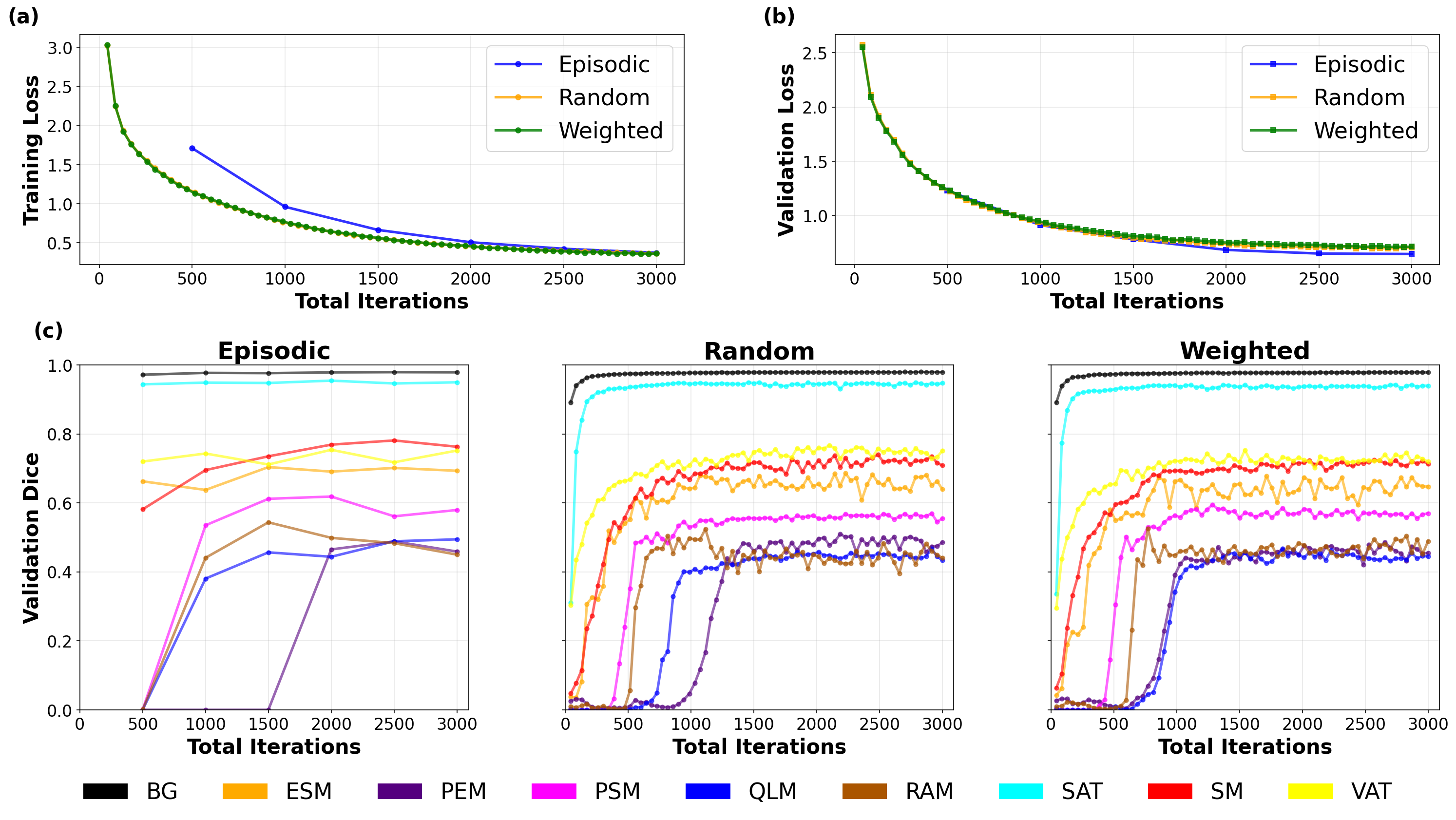}
    \caption{Low-data 10\% fixed 3{,}000 iterations with constant learning rate regime. (a; top--left) training Loss, (b; top--right) validation Loss, (c; bottom) per-class validation Dice performance for episodic (left), random (middle), and weighted (right).}
    \label{fig:experiment_10_fixed}
\end{figure}

\begin{figure}[htp]
    \centering
    \includegraphics[width=\textwidth]{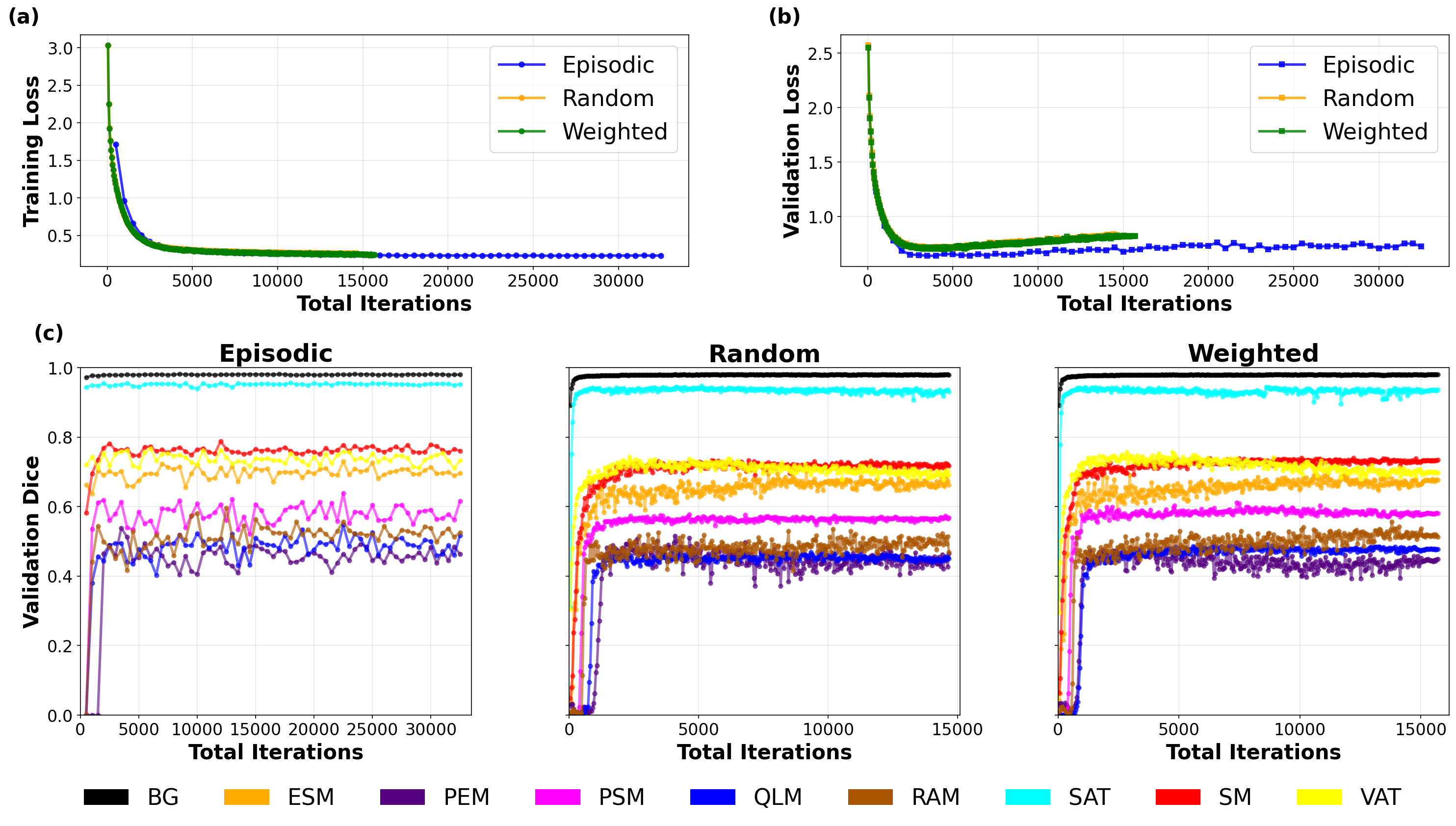}
    \caption{Low-data 10\% iteration-calibrated regime. (a; top--left) training Loss, (b; top--right) validation Loss, (c; bottom) per-class validation Dice performance for episodic (left), random (middle), and weighted (right).}
    \label{fig:experiment_10_calibrated}
\end{figure}

\end{document}